\documentclass{article}

\usepackage{amsmath}
\usepackage{amssymb}
\usepackage{comment}
\usepackage{braket}
\usepackage{graphicx}
\newcommand{\ah}{\hat{a}}
\newcommand{\ahd}{\hat{a}^\dagger}
\newcommand{\bh}{\hat{b}}
\newcommand{\bhd}{\hat{b}^\dagger}
\newcommand{\xh}{\hat{x}}
\newcommand{\ph}{\hat{p}}
\newcommand{\w}{\omega}

\title{Quantum Theory of Distributed-Feedback Parametric Amplifiers and Oscillators}
\author{ Alex O.C. Davis and Alex I. Flint}

\begin{document}

\maketitle
\begin{center}
    
\emph{Centre for Photonics and Photonic Materials, Department of Physics, University of Bath, BA2 7AY, United Kingdom}\\
aocd20@bath.ac.uk

\end{center}



\begin{abstract}
Optical parametric oscillators are among the best-developed quantum light sources, having already been adopted in precision measurement and underpinning various quantum computing and communication paradigms. Meanwhile, progress in photonic structures such as Bragg gratings has enabled distributed feedback oscillators to become widely established as classical laser sources with desirable properties, as well as enabling a new generation of precision optical sensors. Recent work in fabricating and processing photonic structures in nonlinear media opens the path to combining these two programs to realize distributed feedback parametric oscillators. Such devices have great potential as sources of quantum light, especially for squeezed vacuum, a crucial resource state in emerging quantum technologies. We present an analytic and fully quantum-mechanical model of the dynamics of such devices. This approach yields the key properties of these sources, such as the parametric oscillation threshold, intracavity mode, tunability, and quantum statistics (including entanglement) of the output modes. We also discuss the application of these devices as quantum-enhanced sensors. These results underpin future work on a versatile class of next-generation quantum light sources.
\end{abstract}
\section{Introduction}
Optical parametric oscillators (OPOs) are firmly established as important sources of both classical \cite{1449425,canalias2007mirrorless,breunig2011continuous} and quantum \cite{Fabre1990, wang2019quantum,park2024single} light. As classical sources, OPOs are valued for their ability to produce light at new and tunable wavelengths while conserving the coherence properties of the pump(s). In quantum optics, OPOs are unsurpassed in the level of field quadrature squeezing they can demonstrate \cite{PhysRevLett.117.110801}. They have thus found application in quantum-enhanced sensing \cite{PhysRevLett.123.231107,McCuller20}, as well as featuring in proposed applications in quantum information processing \cite{PhysRevLett.101.130501}, state preparation \cite{ra2020non,davis2021conditional}, and communications \cite{PhysRevLett.125.070502}.

Over the years, numerous investigations have sought to combine the advantageous properties of OPOs with the benefits of an all-fiber architecture e.g. \cite{deMatos:04,Zlobina:13,Lamb:13,takahashi2022fiber}. This interest in fibre OPOs (FOPOs) is motivated by a number of considerations, such as high beam quality, high power handling, robustness, compactness, low material losses, and ease of integration with other fiber components. The latter two are especially important for quantum applications, where losses are particularly deleterious to the quality of the source \cite{mcgarry2024microstructured}. To date, however, there has been no reported demonstration of a FOPO acting as a nonclassical light source.

OPOs require feedback and nonlinear interaction. Most fibers are made from amorphous materials, which have no native $\chi^{(2)}$ interaction, so FOPOs depend on $\chi^{(3)}$ processes such as four-wave mixing (FWM). This demands effective phase matching and hence precise dispersion control. For this reason, as well as the higher nonlinearities achievable, solid-core photonic crystal fiber (PCF) is a popular choice of medium. PCF typically consists of an all-silica matrix containing a central core surrounded by a triangular lattice of air holes, the size and spacing of which can be chosen to allow flexible control of the dispersion. Parametric sources based on PCF have been shown to be versatile and effective sources of quantum light \cite{francis2016all}. Previous demonstrations of FOPOs have involved splicing a section of PCF into closed fiber loops to form a ring cavity, providing feedback \cite{deMatos:04,Zlobina:13,Lamb:13,takahashi2022fiber}. While simple and effective, the disadvantage of this approach is that splicing microstructured or ``holey'' fiber is always fairly lossy, with a ring-cavity FOPO normally containing two such splices per round trip. For classical applications, these intracavity losses limit the efficiency and power handling of these devices. They are even more prohibitive for quantum applications, for which the cumulative losses would rapidly degrade the quantum state at the output.

An alternative optical feedback principle that avoids lossy splicing is distributed feedback (DFB), where reflective photonic structures are written directly into the gain medium to form a cavity. DFB is widely employed in the design of various fiber lasers, in which a linear cavity is formed by Bragg gratings (typically with a length of a few millimeters or centimeters) written directly into the fiber. Such grating structures have been employed widely for fiber lasers, with some demonstrations in PCF \cite{Canning:03, Groothoff:05}. The literature contains several studies into DFB OPOs, though most focus on nonlinear crystals rather than fiber. A thorough classical treatment of the singly-resonant, non-degenerate DFB OPO is presented in \cite{Huang:04}. A proposal for a spontaneous parametric down conversion photon pair source based on a singly resonant phase-shift grating cavity in periodically poled nonlinear crystal is given in \cite{Yan:10}. DFB OPOs have been realized in other solid state media \cite{Chiang:02} but not, to the authors' knowledge, in fiber. This can be explained by the fact that conventional PCF, the material of choice for FOPOs, consists only of undoped silica and air. Fiber Bragg grating (FBG) inscription often utilizes colour-centre introduction by UV laser exposure, which requires photosensitivity to the UV. Previous demonstrations of gratings in PCF have, therefore, employed hydrogen loading to induce photosensitivity \cite{Groothoff:03, Canning:03, Groothoff:05}, or used femtosecond lasers \cite{XIANG20189}.

\begin{figure}
\includegraphics[width=\linewidth]{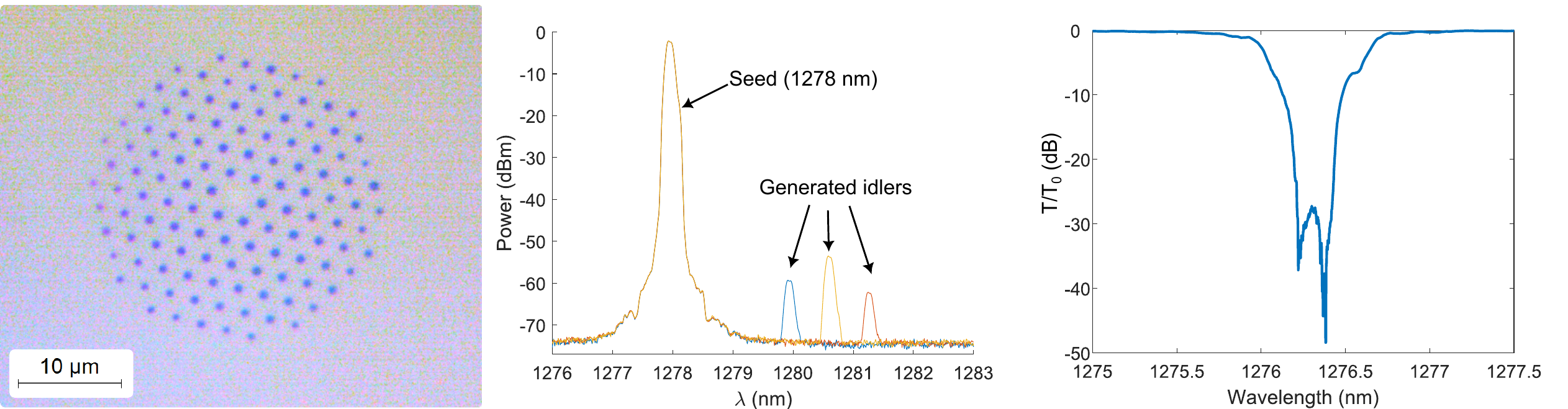}
\centering
\caption{Left: Micrograph of UV-photosensitive germanium-doped PCF, with the Ge-doped region visible as a lighter spot in the fibre core. Centre: Seeded near-degenerate FWM spectra through 10m of this fibre with seed wavelength 1278 nm and pumps $\lambda_p= 1080$ nm and $\lambda_q= 1568$ (blue), $1569$ (orange) and $1570$ nm (red). Right: Normalized transmission spectrum of 50mm Bragg grating inscribed in this fiber.}
\label{fibredetail}
\end{figure}

Recently, we demonstrated the feasibility of direct interferometric UV writing of low-loss, high reflectivity FBGs \cite{davis2023squeezed} into a custom-fabricated PCF with a germanium-doped core (Figure \ref{fibredetail}) \cite{murphy2024tunable}, without hydrogen loading. Our demonstration, achieved inside PCF phase-matched for degenerate FWM around the telecoms O-band, makes a theoretical study into the quantum behavior of DFB FOPOs timely.  

Here we are predominantly concerned with all-fiber devices that can be used for high-gain squeezed vacuum generation. We therefore focus on the FWM process where two non-degenerate pump fields interact to generate a photon pair at the average frequency (see Figure \ref{cavity_relations}, bottom). This FWM pump scheme has been used to generate squeezed vacuum generation in integrated systems \cite{zhang2021squeezed}. We study the case where there is a Bragg grating cavity resonant at the signal wavelength, where degenerate photon pairs are produced. In this scheme the pumps need not interact with the Bragg grating structure and can pass straight through, while the FWM interaction is strongly enhanced due to the resonance at the degenerate signal wavelength. To study this and similar devices, we provide an analytic model of the evolution of the quantum field operators at the signal wavelength in optical waveguides and cavities with both DFB and parametric gain. This allows us to calculate threshold pump powers, required grating properties and field moments (including attainable squeezing levels).

\section{Squeezed light generation in a uniform grating}
We begin by calculating the input-output relations for a FBG of period $\Lambda$ with squeezing in the forward direction. The classical, linear coupled mode equations of the FBG are \cite{erdogan2002fiber}:
\begin{align}
\frac{\partial A}{\partial z}= -i\kappa e^{i\rho z}B \label{coupled1}\\
\frac{\partial B}{\partial z}= i\kappa e^{-i\rho z}A \label{coupled2},
\end{align}
where $A$ and $B$ are the forward and backward propagating field amplitudes, respectively, and $\rho=\beta-K$ is the discrepancy between the effective propagation constant $\beta$ and the Bragg grating wavenumber $K=2\pi/\Lambda$. We promote these to quantum operator equations by making the substitutions
$A\rightarrow\ah$ and 
$B\rightarrow\bh$,
where $\ah$ and $\bh$ are the annihilation operators in the forward- and counter-propagating modes respectively. 

We next evaluate a squeezing term to be added to the evolution of the forward-propagating mode. Since we assume single-pass and continuous wave pumps, peak intensity-dependent nonlinearities such as cross-phase modulation and self-phase modulation can be neglected. In the undepleted pump regime, the forward propagating mode $\ah$ along a section of fibre of length $L$ then evolves under the single-mode squeezing operator: \cite{gerry2023introductory}
\begin{equation}
    \hat{S}(\xi L)=\exp\Big\{\frac{\xi L}{2} \left[\ah^2-\hat{a}^{\dagger 2}\right]\Big\}.
    \label{S_omega}
\end{equation}
We have assumed that the gain parameter $\xi$ is real (without loss of generality) and independent of $z$. For degenerate four-wave mixing, the gain parameter is \cite{agrawal2000nonlinear} $\xi=2\sqrt{P_1P_2}\gamma_{NL}$, where $P_1$ and $P_2$ are the pump powers and $\gamma_{NL}$ is the fibre nonlinearity, typically on the order of $10^{-2}$ W$^{-1}$m$^{-1}$ for PCF. Continuous-wave squeezing measured in practice, such as by balanced homodyne detection, does not usually correspond to monochromatic modes, but rather a mode that is a superposition of two frequencies separated by a demodulation frequency $\Omega$ (usually in the MHz range). Since we will be concerned with resonant structures, where the bandwidths are limited by the cavity linewidth, we will assume that $\Omega\ll\rho c$, such that the squeezed modes can be treated as effectively monochromatic.

To incorporate the effect of the parametric gain on the mode, we take a split-step approach where the modes are incremented by both a beam splitter-like coupling to the backwards-propagating modes (giving rise to Equations \ref{coupled1} and \ref{coupled2} \cite{erdogan2002fiber}) and symplectic coupling between the squeezed mode and its conjugate, which ordinarily gives rise to parametric gain and squeezing. This term takes the form \cite{Lvovsky2015}

\begin{align}
\frac{\partial \ah}{\partial z}= -\xi\ah^\dagger
\end{align}

Combining the coupled-mode and symplectic terms in the forward-propagating mode, Equations \ref{coupled1} and \ref{coupled2} become
\begin{align}
\frac{\partial \ah}{\partial z}&= -i\kappa e^{i\rho z}\bh -\xi\ahd\label{operatorcoupled1}\\
\frac{\partial \bh}{\partial z}&= i\kappa e^{-i\rho z}\ah \label{operatorcoupled2}.
\end{align}
Differentiating Eq. \ref{operatorcoupled1} with respect to $z$ and rearranging gives
\begin{equation}
\frac{\partial^2 \ah}{\partial z^2}+\xi\frac{\partial\ahd}{\partial z}= -i\kappa e^{i\rho z}(i\rho\bh+\frac{\partial\bh}{\partial z} ).
\end{equation}
 Substituting Eqs. \ref{operatorcoupled1} and \ref{operatorcoupled2} into the RHS we obtain
\begin{equation}
\frac{\partial^2 \ah}{\partial z^2}+\xi\frac{\partial\ahd}{\partial z}=i\rho\left(\frac{\partial \ah}{\partial z}+\xi\ahd\right)+\kappa^2\ah \label{evolutioneq}
\end{equation}
We wish to represent the evolution of the field along the length of the fibre as a linear symplectic transformation of the cavity input modes such that $\ah$ and $\ahd$ can be written:
\begin{align}
\ah(z)&=u(z)\ah_0+v(z)\ahd_0+p(z)\bh_L+q(z)\bhd_L \label{sympl}
\end{align}
where $\ah_0=\ah(z=0)$ and $\ah_L=\ah(z=L)$ and similar for $\bh$.
The canonical commutation relation $[\ah(z),\ahd(z)]=1$ implies the condition
$|u(L)|^2+|p(L)|^2-|v(L)|^2-|q(L)|^2=1.$

We can similarly write $b(z)$ as 
\begin{equation}
b(z)=\bar{u}\ah_0+\bar{v}\ahd_0+\bar{p}\bh_L+\bar{q}\bhd_L,
\end{equation}

where $\bar{u},\bar{v},\bar{p},\bar{q}$ can be related to the forward-propagating coefficients $u,v,p,q$ (see Supporting Information). Substituting Equations \ref{sympl} into Eq. \ref{evolutioneq} and collecting terms for the various input operators yields two pairs of coupled differential equations:
\begin{align}
u''+\xi v'^*=i\rho(u'+\xi v^*)+\kappa^2u \\
v''+\xi u'^*=i\rho(v'+\xi u^*)+\kappa^2v
\end{align}
and similarly for $p$ and $q$.  We then insert the \emph{ansatz} solutions $u\sim v^*\sim e^{\lambda z}$, which ultimately leads to a general solution of the form

\begin{equation}
u(z)=U_1e^{\lambda_+z}+U_2e^{-\lambda_+z}+U_3e^{\lambda_-z}+U_4e^{-\lambda_-z},
\label{u_equation}
\end{equation}
where $U_i$ are constants determined by the boundary conditions and \begin{align}
\lambda_\pm^2=\xi^2/2+\kappa^2-\frac{\rho^2}{2} \pm \frac{1}{2}\sqrt{(\rho^2-\xi^2-2\kappa^2)^2+4(\xi^2\rho^2-\kappa^4)}. \label{lambdadefinition}
\end{align}

We also obtain
\begin{equation}
v^*(z)=\gamma_1U_1e^{\lambda_+z}+\gamma_2U_2e^{-\lambda_+z}+\gamma_3U_3e^{\lambda_-z}+\gamma_4U_4e^{-\lambda_-z},
\label{v_equation}
\end{equation}
where 
\begin{equation}
\gamma_i=\frac{-\lambda_i^2+i\rho\lambda_i+\kappa^2}{\lambda_i-i\rho}=\frac{\lambda_i+i\rho}{-\lambda_i^2-i\rho\lambda_i+\kappa^2},
\end{equation}
with $\lambda_i=\{\pm\lambda_\pm\}$.
Applying the boundary condition $\ah(0)=\ah_0$ imposes $u(0)=1$, $v(0)=p(0)=q(0)=0$. Hence  $\sum U_i=1$ and $\sum \gamma_iU_i=0$. The final boundary conditions are imposed by similarly considering the counter-propagating field $\hat{b}(L)$, resulting in a series of linear equations that can be solved to give the field amplitudes in both directions at all points in the grating. Further details on the case with general detuning $\rho \neq 0$ are given in the Supporting Information.

\subsection{Central grating frequency}
These equations simplify considerably if we consider the wavelength where the propagation constant and the grating wavenumber are equal\textendash  the centre wavelength of the grating\textendash  which gives the condition $\rho=0$. This also corresponds to the wavelength where the grating reflectivity and hence cavity finesse is maximised, and is therefore a point of special interest. In this case Eq. \ref{lambdadefinition} reduces to
\begin{equation}
\lambda_\pm=\frac{\pm\xi+\sqrt{\xi^2+4\kappa^2}}{2}.
\end{equation}
and $\gamma_1=\gamma_4=-1, \gamma_2=\gamma_3=1$.

Applying the first set of boundary conditions and noting that $\lambda_\pm=-\lambda_\mp$, Equations \ref{u_equation} and \ref{v_equation} can then be recast in symmetric form
\begin{align}
u(z)&=\frac{\cosh\lambda_+z+\cosh\lambda_-z}{2}+C\cosh\lambda_+z+D\sinh\lambda_+z-C\cosh\lambda_-z+D\sinh\lambda_-z\\
v^*(z)&=\frac{-\sinh\lambda_+z+\sinh\lambda_-z}{2}-C\sinh\lambda_+z-D\cosh\lambda_+z-C\sinh\lambda_-z+D\cosh\lambda_-z
\end{align}

where

\begin{equation}
C=\frac{-1}{2}\frac{\lambda_+^2-\lambda_-^2}{\lambda_+^2+\lambda_-^2+2\lambda_-\lambda_+\cosh(\lambda_++\lambda_-)L}=\frac{1}{2}\frac{\xi\sqrt{4\kappa^2+\xi^2}}{\xi^2-2\kappa^2+2\kappa^2\cosh\sqrt{4\kappa^2+\xi^2} L}
\end{equation}
and
\begin{equation}
D=\frac{-\lambda_+\lambda_-\sinh(\lambda_++\lambda_-)L}{\lambda_+^2+\lambda_-^2+2\lambda_-\lambda_+\cosh(\lambda_++\lambda_-)L}=\frac{-\kappa^2\sinh(\sqrt{4\kappa^2+\xi^2})L}{\xi^2-2\kappa^2+2\kappa^2\cosh\sqrt{4\kappa^2+\xi^2} L}.
\end{equation}
Note that when $\xi=0$, $\lambda_+=\lambda_-=\kappa$ so $C=v(z)=0$ and when $L\rightarrow \infty,~D\rightarrow -1/2$ so $u(z) \rightarrow e^{-\kappa z}$, which recovers the classical solution.

Following similar reasoning for $p$ and $q$, we ultimately find 
\begin{align}
p(z)=R\cosh{\lambda_+z}+S\sinh{\lambda_+z}-R\cosh{\lambda_-z}+S\sinh{\lambda_-z}\\
q^*(z)=-R\sinh{\lambda_+z}-S\cosh{\lambda_+z}-R\sinh{\lambda_-z}+S\cosh{\lambda_-z}.
\end{align}

where

\begin{align}
R=\frac{i\kappa(\lambda_-\sinh\lambda_+L-\lambda_+\sinh\lambda_-L)}{\lambda_+^2+\lambda_-^2+2\lambda_-\lambda_+\cosh(\lambda_++\lambda_-)L} \\
S=\frac{-i\kappa(\lambda_-\cosh\lambda_+L+\lambda_+\cosh\lambda_-L)}{\lambda_+^2+\lambda_-^2+2\lambda_-\lambda_+\cosh(\lambda_++\lambda_-)L}.
\end{align}

\begin{figure}
\includegraphics[width=0.5\linewidth]{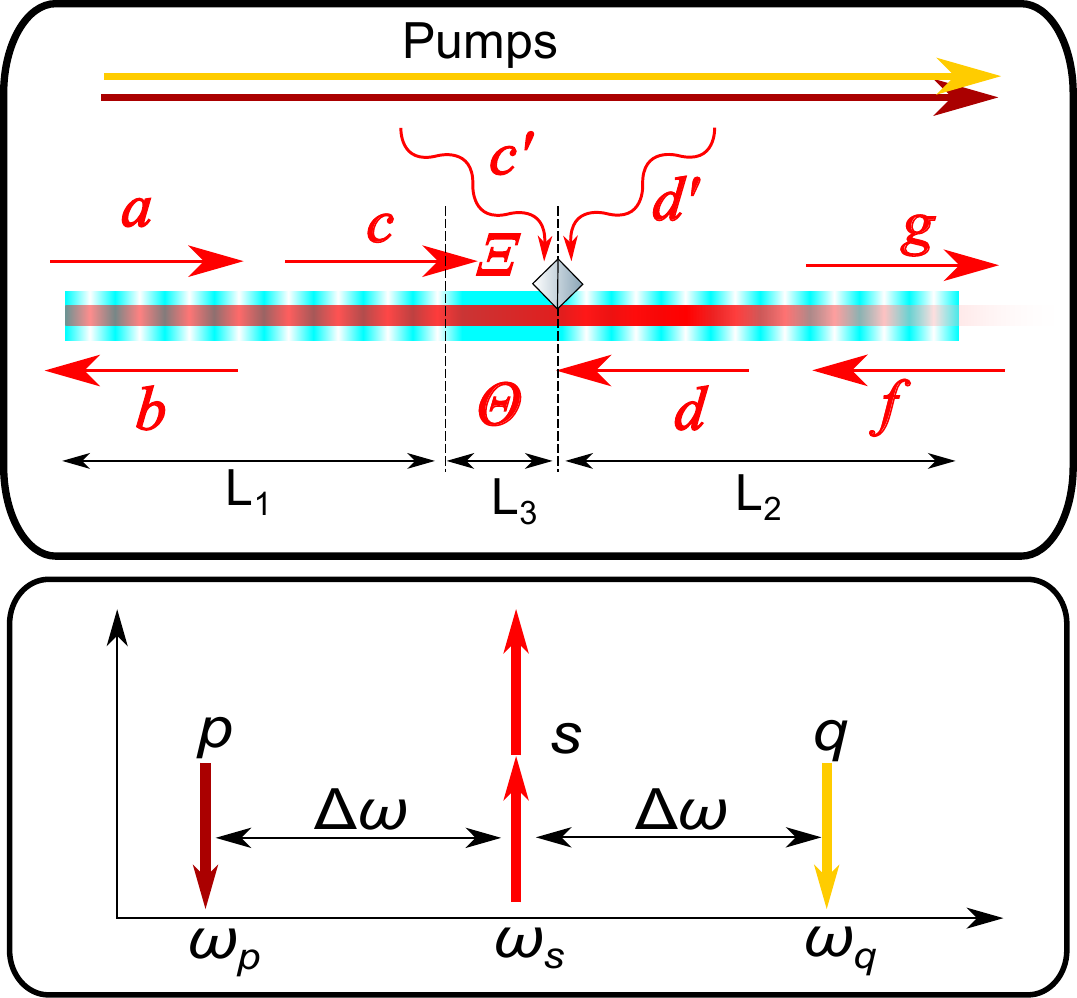}
\centering
\caption{Top: DFB OPO diagram based on dual-pumped four-wave mixing, with key modes labelled, including the noise modes $\hat{c}'$ and $\hat{d}'$. Bottom: Degenerate pumping scheme. Pumps $p$ and $q$ generate photons pairs (squeezing) at their average frequency $\w_s$.}
\label{cavity_relations}
\end{figure}

These results can be considered a quantum generalization of the standard grating reflectivity responses \cite{erdogan2002fiber} to include the simultaneous squeezing transformation.

\section{Optical parametric oscillation}
\subsection{Cavity input-output relations}
We now consider the parametric cavity comprised of two gain-gratings concatenated with an unmodulated region of length $L_3$ between them, imparting a phase shift $\theta$ and an additional squeezing $\xi L_3$ (Figure \ref{cavity_relations}, top). We focus on the cavity field at six key points (see Figure \ref{cavity_relations}): $\ah, \bh$ at $z=0$ (the input/output of the first facet), $\hat{c}, \hat{d}$ (the output modes of the two gratings at the interface) and $\hat{f},\hat{g}$ at $z=L_1+L_2+L_3$ (the input and output modes at the final facet). For clarity, we will presently treat the device as lossless, though we will subsequently investigate the effect of intracavity losses. The modes at general $z$ within the fibre can be determined from these modes using the relations obtained in the previous section.  Using the results of the previous section, we can write the key modes in terms of one another and the functions $u_i,v_i,p_i,q_i$ and $\bar{u}_i,\bar{v}_i,\bar{p}_i,\bar{q}_i$ where $i=\{1,2\}$ indexes the grating (since we do not assume the gratings are the same). We will use the operator-valued vectors $\mathbf{w}=\begin{pmatrix} \hat{w}\\\hat{w}^\dagger\end{pmatrix}$. We can write $\mathbf{c}$ and $\mathbf{d}$ as
\begin{align}
\mathbf{c}=\mathbf{M}_{ca}\mathbf{a}+\mathbf{M}_{cd}\mathbf{\Theta}\mathbf{d}\\
\mathbf{d}=\mathbf{M}_{dc}\mathbf{\Xi}\mathbf{c}+\mathbf{M}_{df}\mathbf{f},
\end{align}
where $\mathbf{M}_{ca}=\begin{pmatrix}u_1(L_1)& v_1(L_1) \\v_1^*(L_1)& u_1^*(L_1)\end{pmatrix}$, $\mathbf{M}_{cd}=\begin{pmatrix}p_1(L_1)& q_1(L_1) \\q_1^*(L_1)& p_1^*(L_1)\end{pmatrix}$, $\mathbf{M}_{dc}=\begin{pmatrix}\bar{u}_2(0)& \bar{v}_2(0) \\ \bar{v}_2^*(0)& \bar{u}_2^*(0)\end{pmatrix}$, $\mathbf{M}_{df}=\begin{pmatrix}\bar{p}_2(0)& \bar{q}_2(0) \\\bar{q}_2^*(0)& \bar{p}_2^*(0)\end{pmatrix}$, $\mathbf{\Theta}=\begin{pmatrix} e^{i\theta}&  0\\ 0& e^{-i\theta}\end{pmatrix}$ and $\mathbf{\Xi}=\begin{pmatrix} e^{i\theta}\cosh \xi L_3&  e^{-i\theta}\sinh \xi L_3\\ e^{i\theta}\sinh \xi L_3& e^{-i\theta}\cosh \xi L_3\end{pmatrix}$, with $L_3$ the length of the unmodulated region.

Substitution of these results into one another gives $\mathbf{c}$ in terms of the input modes $\mathbf{a}$ and $\mathbf{f}$:
\begin{align}
\mathbf{c}=\bar{\Delta}(\mathbf{M}_{ca}\mathbf{a}+\mathbf{M}_{cd}\mathbf{\Theta}\mathbf{M}_{df}\mathbf{f}),
\end{align}
where $\bar{\Delta}=(\mathbb{I}_2-\mathbf{M}_{cd}\mathbf{\Theta}\mathbf{M}_{dc}\mathbf{\Xi})^{-1}$ and $\mathbb{I}_2$ is the $2 \times 2$ identity matrix. This expression diverges when 
\begin{equation}
\det(\mathbb{I}_2-\mathbf{M}_{cd}\mathbf{\Theta}\mathbf{M}_{dc}\mathbf{\Xi})=0,
\end{equation}
 which gives the condition for the parametric oscillation pump threshold. We can similarly express $\mathbf{d}$:
\begin{align}
\mathbf{d}=\Delta(\mathbf{M}_{df}\mathbf{f}+\mathbf{M}_{dc}\mathbf{\Xi}\mathbf{M}_{ca}\mathbf{a})
\end{align}
where $\Delta=(\mathbb{I}_2-\mathbf{M}_{dc}\mathbf{\Xi}\mathbf{M}_{cd}\mathbf{\Theta})^{-1}$. The cavity outputs can then be expressed in terms of the input functions:
\begin{align}
\mathbf{b}=\mathbf{M}_{ba}\mathbf{a}+\mathbf{M}_{bd}\mathbf{\Theta}\mathbf{d}\\
\mathbf{g}=\mathbf{M}_{gf}\mathbf{f}+\mathbf{M}_{gc}\mathbf{\Xi}\mathbf{c}
\end{align}
where $\mathbf{M}_{ba}=\begin{pmatrix}\bar{u}_1(0)& \bar{v}_1(0) \\ \bar{v}_1^*(0)& \bar{u}_1^*(0)\end{pmatrix}$, $\mathbf{M}_{bd}=\begin{pmatrix}\bar{p}_1(0)& \bar{q}_1(0) \\\bar{q}_1^*(0)& \bar{p}_1^*(0)\end{pmatrix}$, $\mathbf{M}_{gf}=\begin{pmatrix}p_2(L_2)& q_2(L_2) \\q_2^*(L_2)& p_2^*(L_2)\end{pmatrix}$ and $\mathbf{M}_{gc}=\begin{pmatrix}u_2(L_2)& v_2(L_2) \\v_2^*(L_2)& u_2^*(L_2)\end{pmatrix}$. This lets us write the transformation due to the gain-gratings in terms of a beamsplitter-like operation:
\begin{equation}
\mathbf{b}=\mathbf{R}_{ba}\mathbf{a}+\mathbf{T}_{bf}\mathbf{f},
\label{outputcoupling}
\end{equation}
where
\begin{align}
\mathbf{R}_{ba}&=\mathbf{M}_{ba}+\mathbf{M}_{bd}\mathbf{\Theta}\Delta\mathbf{M}_{dc}\mathbf{\Xi}\mathbf{M}_{ca}\\
\mathbf{T}_{bf}&=\mathbf{M}_{bd}\mathbf{\Theta}\Delta\mathbf{M}_{df}
\label{rant}
\end{align}

\subsection{Squeezing and output quantum correlations}
We can reformulate the cavity input-output relations in terms of the quadrature operators for a mode $\mathbf{w}$, $\xh_w=(\hat{w}+\hat{w}^\dagger)/2$ and $\ph_w=-i(\hat{w}-\hat{w}^\dagger)/2$, expressed in operator-valued vector form:
\begin{equation}
\mathbf{x}_w=\begin{pmatrix} \hat{x}_w \\ \hat{p}_w \end{pmatrix}=\mathbf{Q}\mathbf{w},
\end{equation}
where $\mathbf{Q}=\frac{1}{2}\begin{pmatrix}1&1\\-i&i\end{pmatrix} $ and $\mathbf{Q}^{-1}=\begin{pmatrix}1&i \\1 &-i\end{pmatrix}$.

Hence we can write the beamsplitter transformation of the quadrature operators
\begin{equation}
\mathbf{x}_b=\mathbf{\tilde{R}}_{ba}\mathbf{x}_a+\mathbf{\tilde{T}}_{bf}\mathbf{x}_f,
\end{equation}
where
\begin{equation}
\mathbf{\tilde{R}}_{ba}=\begin{pmatrix} R_{xx}& R_{xp}\\R_{px}&R_{pp}\end{pmatrix}=\mathbf{Q}\mathbf{R}_{ba}\mathbf{Q}^{-1},
\end{equation}
and similarly for $\mathbf{\tilde{T}}_{bf}$.
We can then express the quadrature variables of the output in terms of those of the input:
\begin{align}
\xh_b=R_{xx}\xh_a+R_{xp}\ph_a+T_{xx}\xh_f+T_{xp}\ph_f \label{x_b}\\
\ph_b=R_{px}\xh_a+R_{pp}\ph_a+T_{px}\xh_f+T_{pp}\ph_f.
\end{align}
Assuming the injected modes $\ah$ and $\hat{f}$ are in the vacuum state, then $\braket{\xh_i}=\braket{\ph_i}=\braket{\xh_i\ph_j}=\braket{\ph_i\xh_j}=0$ and $\braket{\xh_i^2}=\braket{\ph_i^2}=1/2$ for $i,j=a,f$. Hence $\braket{\xh_b}=\braket{\ph_b}=0$, and we can write the expectation values of the variances as
\begin{align}
\braket{\xh_b^2}=(R_{xx}^2+R_{xp}^2+T_{xx}^2+T_{xp}^2)/2 \\
\braket{\ph_b^2}=(R_{px}^2+R_{pp}^2+T_{px}^2+T_{pp}^2)/2. \label{p_var}
\end{align}
The (anti-)squeezing in dB is then given by $-10\log_{10}\{2\braket{\xh^2}\}~~(10\log_{10}\{2\braket{\ph^2}\})$ and the photon number expectation given by $\braket{\hat{n}_b}=(\braket{\xh_b^2}+\braket{\ph_b^2}-1)/2$. By repeating this procedure with the second output mode $\hat{g}$, we can fully quantify the correlations between the two outputs of the device by the covariance matrix of the two pairs of field quadratures. In general, when both output couplers are comparably transmissive, the two modes will exhibit correlations in their quadratures. If the minimal eigenvalue of the joint covariance matrix is less than unity, then the modes are entangled with one another (see Figure \ref{phaseshiftgratingfig} b)). 

\subsection{Intracavity loss}
We can model intracavity losses in the system with a beam splitter of reflectivity $r$ placed in the centre of the cavity, coupling the modes $\hat{d}$ and $\Xi\hat{c}$ with two injected modes $\hat{c}'$ and $\hat{d}'$ (see Fig. \ref{cavity_relations}). In this case Equation \ref{outputcoupling} becomes 
\begin{equation}
\mathbf{b}=\mathbf{R}'_{ba}\mathbf{a}+\mathbf{T}'_{bf}\mathbf{f}+r(\Gamma_1\mathbf{c}'+\Gamma_2\mathbf{d}'),
\end{equation}

where $\mathbf{R}'_{ba}$ and $\mathbf{T}'_{bf}$ are given by Equations \ref{rant} with $\Delta \rightarrow\Delta'=(\mathbb{I}_2-(1-|r|^2)\mathbf{M}_{cd}\mathbf{\Theta}\mathbf{M}_{dc}\mathbf{\Xi})^{-1}$ and the loss coupling matrices are given by
\begin{align}
\Gamma_1&= \mathbf{M}_{bd}\mathbf{\Theta}\Delta'\mathbf{M}_{dc}\\
\Gamma_2&=\mathbf{M}_{bd}\mathbf{\Theta}\Delta'\mathbf{M}_{dc}\Xi\mathbf{M}_{cd}\mathbf{\Theta}
\end{align}

Following the procedure of Equations \ref{x_b}-\ref{p_var} with the inclusion of these loss modes (which we also assume to be initialised in the vacuum state) the loss-dependent squeezing can be found. The minimal eigenvalue of the covariance matrix for a phase shift grating cavity with equal couplers, with varying degrees of pump power and intracavity loss, is shown in Figure \ref{phaseshiftgratingfig} b). Similar to the more familiar case of a Fabry-Perot OPO with hard reflectors, we find that the strongest squeezing is found close to threshold at low loss, but at these pump powers squeezing is most fragile to increasing loss.

Propagation losses in similar Ge-doped PCF were measured by cutback measurements at 0.03 and 0.05 dB m$^{-1}$ at 920 nm and 1550 nm respectively \cite{murphy2024tunable}, suggesting that for devices a few tens of centimeters long operating in the near infrared, finesses on the order of 100 are achievable before waveguide losses become dominant.

\subsection{Fabry-Perot cavity}

\begin{figure}
    \includegraphics[width=\linewidth]{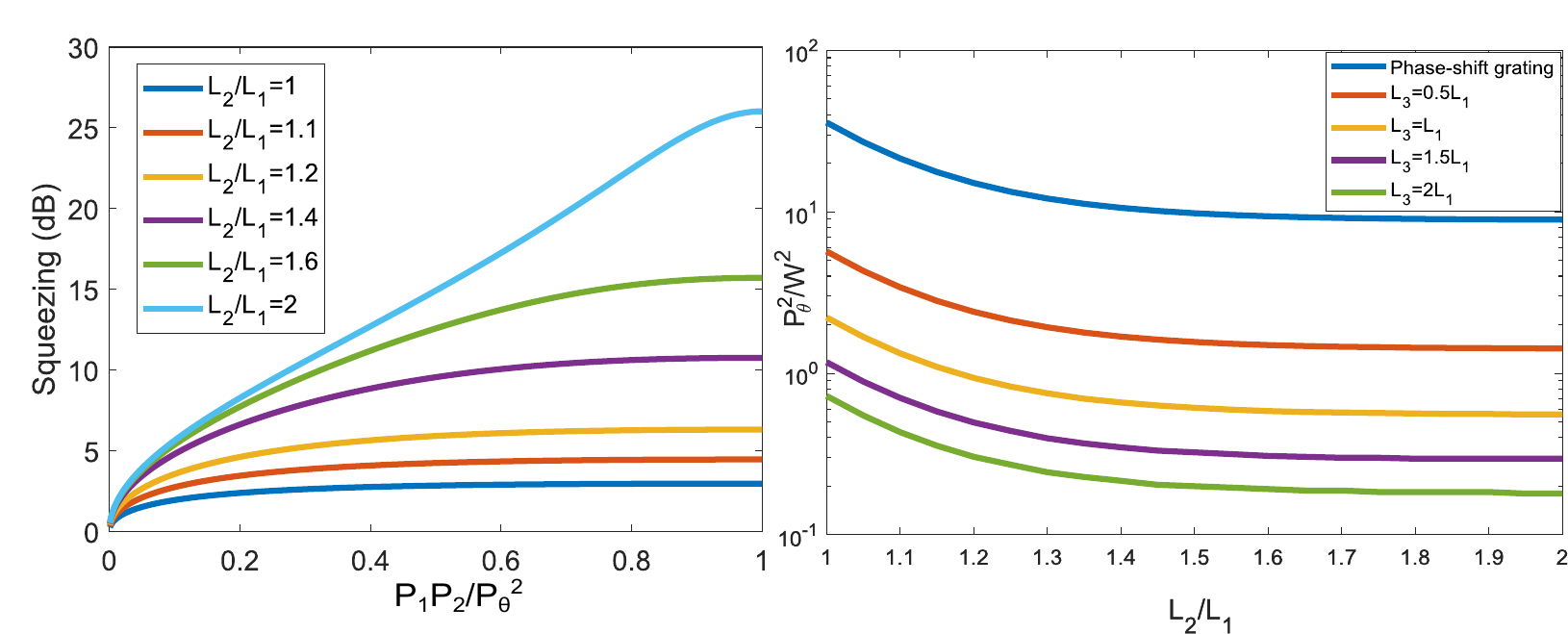}
    \centering
    \caption{Left: Dependence of sub-threshold squeezing in the $\hat{b}$ mode with pump power and relative grating length for a Fabry-Perot cavity, with $\kappa L_1=3$ and $L_3=4L_1$. Right: Semilogarithmic plot of the oscillation threshold pump power for various combinations of $L_2,~L_3$. Here $\kappa L_1=3$, $L_1=0.1$m and $\gamma_{NL}=0.25$ W$^{-1}$km$^{-1}$.}
    \label{SVP}
\end{figure}

A Fabry-Perot cavity is formed when a sizeable section of unmodulated material is left between the two gratings, permitting a comb of resonant modes. When the length of this section $L_3\gg1/\kappa$, most of the parametric interaction takes place in this region, and the behavior of the device tends to that of a conventional linear-cavity OPO with hard reflectors.
Figure \ref{SVP} (left) shows the variation of the squeezing with pump power below threshold ($P_1P_2=P_\theta^2$) in various Fabry-Perot cavities. Here $\kappa L_1=3$ (corresponding to a grating reflectivity of around 99\%, $L_3=4L_1$ and $L_2$ varies from $L_1$ to $2L_1$. When $L_1=L_2$, the squeezing at one output tends to a maximum value of around 3dB, a limit imposed by the effective 50\% losses through the other coupler. As $L_2/L_1$ is increased, the proportion of the light coupled out through the first coupler increases so the squeezing grows. When the loss through the second grating into mode $\hat{g}$ approaches zero, the squeezing at high pump powers begins to diverge. For realistic values of $\gamma_{NL}=2.5\times 10^{-2}$ W$^{-1}$m$^{-1}$ and $L_1=50$ mm, the threshold pump powers for these devices vary from 0.21 W$^2$ when $L_2=2L_1$ to 0.84 W$^2$ when $L_2=L_1$. Figure \ref{SVP} (right) shows the dependence of the oscillation threshold pump power $P_\theta$ on the relative lengths of $L_1,~L_2,~L_3$. The lowest threshold is achieved when $L_2$ and $L_3$ are long, giving the strongest optical feedback and the greatest interaction length respectively.

\subsection{Phase-shift grating cavity}
A second cavity architecture, possible only in DFB systems, is the phase-shift grating cavity. This consists of a single continuous grating interrupted by a $\pi/2$ phase shift in the middle, supporting a single sharp resonant mode within the stop band of the grating. This is commonly used in doped fibers for DFB fiber lasers \cite{Groothoff:05, maniewski2024tunable}, as well as in DFB Raman \cite{Loranger:17} and Brillouin lasers \cite{Abedin:12}. However, such structures have not been employed in FOPOs. This design is desirable when a single, stable, narrowband emission peak is required, but this comes at the cost of the long interaction region with resonant field enhancement in the Fabry-Perot geometry.

\begin{figure}
\centering
    \includegraphics[width=\linewidth]{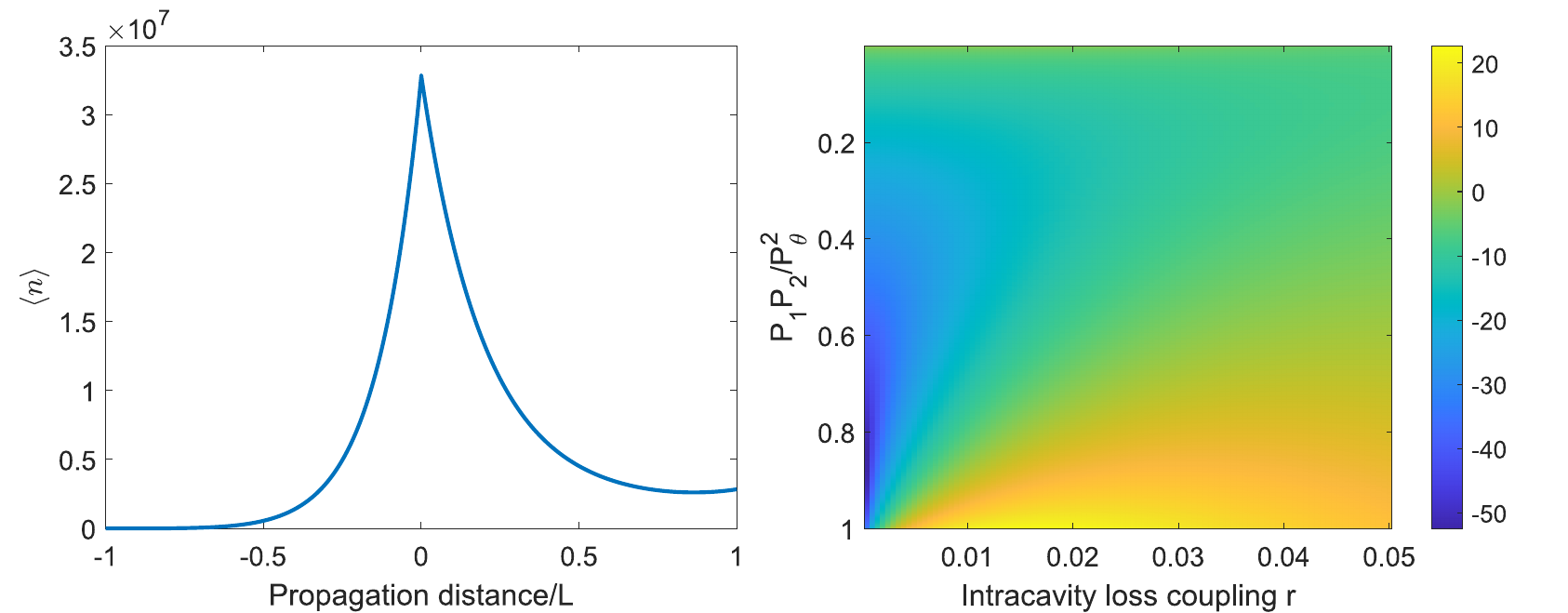}
    \caption{LEFT: Expectation value for the photon number per mode volume in the forward propagation direction along the length of a phase-shift grating with $L_1=L_2=L$, $\kappa L=1.5$ and $P_1P_2=0.9P_\theta^2$ with the defect located at $z=0$. The double-exponential mode envelope characteristic of phase-shift gratings is modified by the gain. Note that in the final section of the device the intensity is increasing again. RIGHT: Suppression of the quadrature noise ($10\log_{10}\langle \xh^2\rangle$) of the most squeezed supermode of the two outputs of a phase shift grating with $\kappa L_1=\kappa L_2=3$, for varying levels of loss and pump power.}
    \label{phaseshiftgratingfig}
\end{figure}

We can model $\pi/2$ phase shift FOPOs by considering the two sections on either side of the interruption as separate gain-gratings, and setting $\theta=\pi/2$ and $L_3\approx 0$. In a linear-optical phase-shift grating cavity, the resonant mode has a characteristic double-exponential shape with a cusp at the defect and boundary conditions determined by the external driving field. As we have seen, when parametric gain is added to the calculation, these exponential decay curves are modified into summed hyperbolic functions. The mean photon number (intensity) along the $z$ direction shows a profile that is superficially similar to the double-exponential solution, but shows the effect of gain along the $z$ axis (Figure \ref{phaseshiftgratingfig}). A qualitative difference with the linear optical case is that the decay curve down-pump of the defect reaches a local minimum inside the grating before increasing again up to the output facet. 

\subsection{As quantum-enhanced fiber sensors}
We have discussed how DFB-OPOs can be useful as quantum light sources, by bringing the various advantages of DFB lasers into the domain of quantum light sources. Additionally, DFB-OPOs have potential for application as quantum-enhanced sensors. Classically, DFB fiber cavities are widely used in sensing applications due to their sub-wavelength scale strain sensitivity and robustness to external electrical and radiation environments. In particular, resonant fiber cavities are used as interferometric sensors due to the strong sensitivity of the phase at the output to small variations in the intracavity length induced by strain or temperature variations \cite{GIURGIUTIU2016249}.

DFB-OPOs present an opportunity to introduce quantum advantage to such sensors. As with all interferometric phase estimators, the sensitivity is fundamentally limited by the phase noise of the probe signal, which is determined by its quantum state. With a passive cavity, phase sensitivity is limited by noise in the phase quadrature of the signal passing through the sensor, which for a coherent state is constant with $\theta$ and scales as $\Delta\theta^2\sim 1/N$, where $N$ is the number of photons. When inline squeezing is employed the phase noise can be squeezed around the values of $\theta$ where the interferometer is set, giving a more precise measurement around this set of values.

\begin{figure}
    \includegraphics[width=0.8\linewidth]{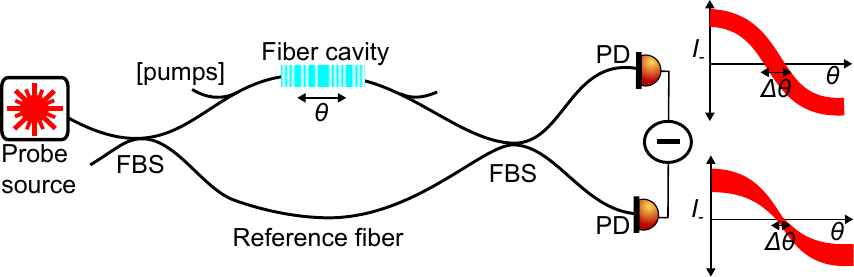}
    \centering
    \caption{Schematic diagram of a balanced interferometric fiber sensor using the round-trip phase of a fiber cavity to measure an applied strain. With a coherent-state probe (top right), phase sensitivity is limited by noise in the phase quadrature of the signal passing through the sensor. When inline squeezing is employed (bottom right) the phase noise can be squeezed around the values of $\theta$ where the interferometer is set, giving a more precise measurement. FBS- fibre beam splitter. PD- photodiode. $I_-$- difference photocurrent.}
\end{figure}

\section{Conclusion}
We have presented an analytic and fully quantum-mechanical model for degenerate photon pair generation (single-mode vacuum squeezing) in one-dimensional DFB devices. Our model allows us to calculate threshold powers, the moments of the fields at the two outputs and intracavity intensities, and accounts for spectral detuning from the Bragg resonance. These results can be straightforwardly adapted to consider externally seeded fields and/or operation in the classical regime. Our model can accommodate small losses by modifying the model parameters. We are motivated by the expected development of DFB FOPOs as quantum light sources, which have potential applications in quantum technologies, especially quantum-enhanced sensing and quantum networking. However, our model is general and suitable for other platforms including integrated waveguides and photonic chips, supporting degenerate pair production either by FWM or spontaneous parametric down conversion.

\medskip
\textbf{Supporting Information} \par 
Supporting Information and simulation code are available from the authors.

\medskip
\textbf{Acknowledgements} \par 
AOCD acknowledges support from UKRI/EPSRC through grant reference  EP/W028336/1 ``PHOCIS: A Photonic Crystal Integrated Squeezer". We thank Leah Murphy and Alex McMillan for figure credits and Peter Horak for useful comments. We declare no conflicts of interest. 

\medskip

%
\bibliographystyle{unsrt}
\bibliography{main}

\end{document}